\documentclass{ws-p8-50x6-00}

\usepackage{times}
\newcommand{\Preprint}[1]{\mbox{}\hfill{\normalsize\sf #1}\\[2cm]}

\begin{document}

%% preprint No. SNUTP-98-148

\title{\Preprint{\vbox{\hbox{SNUTP-98-148}\hbox{nucl-th/9901021}}}
$\phi$-meson photoproduction and strangeness of the
nucleon%
\footnote{Talk presented at the 1998 YITP Workshop on QCD and Hadron
Physics, Oct. 14--16, 1998, Kyoto, Japan.}}

\author{Yongseok Oh%
\footnote{E\lowercase{-mail address: yoh@phya.snu.ac.kr}}}

\address{
Center for Theoretical Physics, Seoul National University,
     Seoul 151-742, Korea}

\maketitle

\abstracts{
The polarization observables in $\phi$ photoproduction are
suggested for probing the nucleon strangeness.
Based on models for $\phi$ photoproduction, some double
polarization observables are shown to be very sensitive to the
strangeness content of the proton because of the different spin
structures of the amplitudes associated with different mechanisms.}

The possible existence of hidden strangeness in the nucleon has been one
of the most controversial problems in hadron physics.
Some experimental results have been interpreted as signatures of hidden
strangeness of the nucleon\cite{EMC,EGK}, but it has been also claimed
that these experiments could be understood with little or null
strangeness in the nucleon\cite{ASL}.

It will be interesting, therefore, to study other processes that might be
related directly to the hidden nucleon strangeness\cite{MDDP}.
One of them is to use $\phi$ production from the proton.
Since the $\phi$ is nearly pure $s \bar s$ state, its
coupling to the proton is suppressed by the OZI rule.
Then the idea is that studying the strange sea quark contribution through
the OZI evasion processes may give informations about the hidden strangeness of
the nucleon.

\begin{figure}[t]
\centerline{\epsfig{file=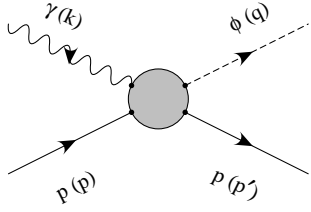, width=0.35\hsize}}
\caption{Kinematics for $\phi$ meson photoproduction from the proton,
$\gamma p \to \phi p$.}
\label{fig01}
\end{figure}

In this work, we discuss $\phi$ photoproduction process from the proton
targets (Fig. \ref{fig01}).
We define $k$, $q$, $p$, and $p'$ as the four-momenta of the photon,
$\phi$-meson, target proton, and recoiled proton, respectively, and
$t = (p-p')^2$ and $W^2 = (p+k)^2$. The scattering angle
in c.m. frame is denoted by $\theta$.
In this case, the main contribution to the cross section is from
the diffractive production process of vector-meson dominance (VDM), and the
one-pion-exchange (OPE) process comes in as a correction to VDM\cite{VDM}.
This OPE process is possible because of the $\phi$-$\pi$-$\gamma$
coupling, and represents the contributions from the (small) non-strange
quark components of the $\phi$-meson.

In addition to these processes, the contribution from the hidden
strangeness of the proton arises through the direct knockout process.
For this purpose, we write the proton wave function in Fock space
as\cite{henley,TOY}
\begin{eqnarray}
|p\rangle = A|[uud]^{1/2}\rangle + \sum_{j_{s\bar s}=0,1; j_c}
b_{j_{s\bar s}} \ [[[uud]^{1/2} \otimes [{\bf L}]]^{j_c} \otimes [s\bar
s]^{j_{s\bar s}} ]^{1/2} \rangle,
\end{eqnarray}
where the superscripts denote the spin of each cluster and $(b_0,b_1)$
correspond to the amplitudes of the $s\bar s$ cluster with spin 0 and 1,
respectively. Then the strangeness admixture of the proton is $B^2 =
b_0^2 + b_1^2$. The symbol $\otimes$ represents vector addition of the
cluster spin and the orbital angular momentum ${\bf L}$ ($\ell=1$).
The purpose of this study is to extract an information about $B^2$, if it
is nonzero, through $\phi$ photoproduction from the proton targets.

\begin{figure}[t]
\centerline{\epsfig{file=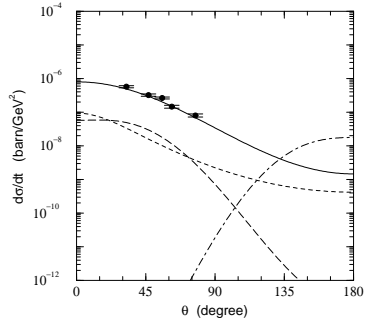, width=0.40\hsize}}
\caption{Unpolarized differential cross section $d\sigma/dt$ of $\phi$
photoproduction at $W = 2.155$ GeV. The solid, dotted, dashed, and
dot-dashed lines give the cross section of VDM, OPE, $s\bar s$-knockout,
and $uud$-knockout, respectively, with strangeness admixture $B^2=1$\%
assuming $|b_0| = |b_1|$.
The experimental data are from Ref. 8.}
\label{fig02}
\end{figure}

In Ref. 6, Henley {\it et al.\/} calculated the contribution
of the knockout process in $\phi$ electroproduction cross section using
nonrelativistic harmonic oscillator quark model and
found it comparable to that of VDM with an assumption of $B^2 = 10$-$20$\%.
This work was improved by Titov {\it et al.\/}\cite{TOY} with the use of
a relativistic harmonic oscillator quark model and leads to the
conclusion that a theoretical upper bound of $B^2$ would be less than 5\%.
In Fig. \ref{fig02}, we give our results for the cross section for $\phi$
photoproduction.
As can be seen from Fig. \ref{fig02}, it is not easy to distinguish the
mechanisms from the cross section measurements because their respective
contributions have similar dependence on the momentum transfer.

\begin{figure}[t]
\centerline{\epsfig{file=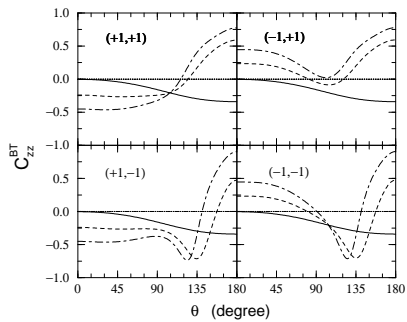, width=0.45\hsize}}
\vskip -24pt
\caption{The double spin asymmetry $C_{zz}^{\rm BT}$ at $W=2.155$ GeV
with $B^2 = 0$\% (solid lines), $0.25$\% (dashed lines), and 1\%
(dot-dashed lines) assuming $|b_0| = |b_1|$.
The phases $(\eta_0,\eta_1)$ of $(b_0, b_1)$ are given explicitly in
each graph.}
\label{fig03}
\end{figure}

It is, therefore, required to find other quantities which are very
sensitive to the nucleon strangeness and we found that some double
polarization observables in $\phi$ photoproduction could be very useful
tools to study the nucleon strangeness\cite{TOYpol}.
Having the helicity amplitudes of each process, we can compute various
polarization observables using the formalism developed, {\it e.g.\/}, in
Ref. 10.
In Fig. \ref{fig03}, as a typical example, we show our results
for the double polarization observable $C^{\rm BT}_{zz}$ where both the
photon and the target proton are polarized along the direction of the
photon momentum;
\begin{equation}
C^{\rm BT}_{zz} \equiv
\frac{d\sigma_{\uparrow\uparrow} - d\sigma_{\uparrow\downarrow}}
{d\sigma_{\uparrow\uparrow} + d\sigma_{\uparrow\downarrow}},
\end{equation}
where $d\sigma$ represents $d\sigma/dt$ and the arrows denote the photon
beam and target proton helicities.
One can find that this quantity is very sensitive to the hidden nucleon
strangeness $B^2$. At forward scattering angles, $C^{\rm BT}_{zz}$
approaches zero with $B^2=0$, but even with $B^2 = 1$\% its magnitude is
as large as 0.45.

This remarkable sensitivity of $C^{\rm BT}_{zz}$ on $B^2$ comes
from the different spin structures of the associated amplitudes.
At forward scattering angle limit, the helicity amplitudes read
\begin{eqnarray}
&& H^{\rm VDM} \to -i M_0^{\rm VDM}
\delta_{\lambda_f \, \lambda_i} \delta_{\lambda_\phi\, \lambda_\gamma},
\qquad
H^{\rm OPE} \to - M_0^{\rm OPE} (2\lambda_i \lambda_\gamma)
\delta_{\lambda_f \, \lambda_i} \delta_{\lambda_\phi\, \lambda_\gamma},
\nonumber \\ &&
H^{s\bar s} \to - i M_0^{s \bar s} (2\lambda_i \lambda_\gamma)
\delta_{\lambda_f \, \lambda_i} \delta_{\lambda_\phi\, \lambda_\gamma},
\end{eqnarray}
where the helicities of the target and recoiled proton, photon beams, and
$\phi$-meson are denoted respectively by $\lambda_{i,f,\gamma,\phi}$. The
relevant amplitudes of each process are given by $M_0^{\rm VDM}$, etc.
The $uud$-knockout is suppressed at forward scattering angles and we can
ignore it in this region.
Then the above analysis gives
\begin{equation}
C^{\rm BT}_{zz} \simeq
- 2 \eta_0 \sqrt{\sigma^{s\bar s}/\sigma^{\rm VDM}},
\end{equation}
where $\eta_0$ is the phase of $b_0$.
This shows that the above quantity is very sensitive to $\sigma^{s\bar s}$
as verified by numerical calculations given in Fig. \ref{fig03}.

As a summary, we found that some double polarization observables in
$\phi$ photoproduction could be very useful for investigating the hidden
strangeness of the nucleon. The optimal range of the initial photon
energy to measure $B^2$ would be around 2-3 GeV, which can be reached by the
current electron facilities, and several experiments have been
proposed at JLab and RCNP to measure these quantities.

\bigskip
This work was done in close collaboration with A. I. Titov, S. N. Yang, and
T. Morii. I am grateful to them for valuable discussions. I also thank
D.-P. Min and C.-R. Ji for encouragements. This work was supported in
part by the Korea Science and Engineering Foundation through the CTP of
Seoul National University.

\end{document}